\documentclass{PoS}
\usepackage{subcaption}

\newcommand{\ra}{\rightarrow}

\newcommand{\is}{ &\! =\! & }
\newcommand{\nonum}{\nonumber \\[1.5mm]}

\newcommand{\inv}{^{-1}}
\newcommand{\ie}{{\it i.e.\ }}

\renewcommand{\th}{{\theta}}

\newcommand{\lb}{\lambda}

\newcommand{\Gm}{\Gamma}
\newcommand{\gm}{\gamma}
\newcommand{\ph}{\phi}

\newcommand{\dd}{{\partial}}

\title{Critical behavior of the hopping expansion from the  Functional Renormalization
Group}

\ShortTitle{Criticality in hopping expansion from LPA}

\author{\speaker{R. Banerjee}\\
	Pittsburgh Particle Physics, Astrophysics, and Cosmology Center,\\
	 Department of Physics and Astronomy,\\
        University of Pittsburgh, Pittsburgh, PA 15260, U.S.A\\
        E-mail: \email{rub18@pitt.edu}}

\abstract{A lattice version of the widely used Functional
Renormalization Group (FRG) for the Legendre effective action is
solved - in principle exactly - in terms of graph rules for
the linked cluster expansion. Conversely, the FRG induces
nonlinear flow equations governing suitable
resummations of the graph expansion. The (finite) radius of
convergence determining criticality can then be efficiently
computed as the unstable manifold of a Gaussian or non-Gaussian
fixed point of the FRG flow. The correspondence is tested on the
critical line of the L\"{u}scher-Weisz solution
of the $\phi^4_4$ theory and its $\phi_3^4$ counterpart.}

\FullConference{The 36th Annual International Symposium on Lattice Field Theory
  - LATTICE2018\\22-28 July, 2018\\
Michigan State University, East Lansing, Michigan, USA.}

\begin{document}

\section{Introduction and Motivation}

The functional renormalization group (FRG) has become one of the most widely
and fruitfully used techniques in quantum many body physics, and is now 
applied to areas as diverse as quantum gravity, particle physics,
and solid state physics \cite{FRGbooks}. The FRG is a reformulation of
quantum field theory  that focusses on the non-linear response of functionals
to a scale dependent mode modulation introduced  by replacing the bare action
$S[\chi]$ with $S[\chi]+\frac{1}{2} \chi\cdot R_k\cdot \chi$
in the functional integral. The regulator kernel $R_k$ supresses low
energy modes and vanishes at $k=0$, such that the scale $k$ smoothly
interpolates between the bare theory and the renormalized theory. Modern
formulations focus on the Legendre effective action $\Gm_k$, whose flow satisfies
\begin{equation}
\label{FRG1}
\dd_k\Gm_k[\phi]=\frac{1}{2}{\rm Tr}\big\{\dd_k R_k[\Gm_k^{(2)}+R_k]^{-1} \big\}\,.
\end{equation}
The versatility of the flow equation (\ref{FRG1}) is partly due to its kinematical
nature; dynamical information is injected solely through initial conditions. As a
consequence, fully non-perturbative results require some such initial conditions. 
An especially good choice are ultralocal initial conditions 
as they can, in a lattice formulation, be computed exactly
from single site integrals \cite{Dupuis-Machado}. A solution of (\ref{FRG1}) 
with such initial data, if feasible, will emulate a linked cluster or hopping 
expansion but with a scale dependent long-ranged interaction 
\begin{equation} 
\label{i2}
S[\chi] = \sum_x s[\chi_x] + \frac{\kappa}{2} 
\sum_{x,y} \chi_y \ell_{xy}(k) \chi_x\,. 
\end{equation}
For definiteness we consider here a self-interacting, one-component, 
scalar field theory on a $D$-dimensional hypercubic lattice
(identified with $\mathbb{Z}^D$) in a dimensionless formulation. 
Then, $s:\mathbb{R} \rightarrow \mathbb{R}$ is a real even function bounded from below
that collects all terms referring to a single site. The hopping parameter $\kappa>0$
is a dimensionless combination of the original mass, the coupling 
parameters, and the lattice spacing. A fundamental lattice action
has the form (\ref{i2}) with a $k$-independent $\ell_{xy}$ that
connects only nearest neighbors. In order to relate (\ref{FRG1})
to a hopping expansion we take $\kappa$ itself as the control parameter
and replace (\ref{FRG1}) by 
\begin{eqnarray}
\label{i3}
\partial_{\kappa} \Gamma_{\kappa} = 
\frac{1}{2} \sum_{x,y} \ell_{xy} \big[ \Gamma_{\kappa}^{(2)} + \kappa \ell 
\big]_{xy}^{-1} \,,\quad 
\Gamma_{\kappa}[\phi] = \Gamma_0[\phi] + \sum_{l \geq 2} \kappa^{l} 
\Gamma_{l}[\phi]\,.
\end{eqnarray}
Here $\Gamma_0[\phi] = \sum_x \gamma(\phi_x)$, where $\gamma$ and its derivatives $\gm_n$
are computable at a single site $x$ from $s$ only. The $O(\kappa)$ 
term vanishes, $\Gamma_2[\phi] 
= -\frac{1}{4} \sum_{x,y} (\ell_{xy})^2 \gm_2(\phi_x)\inv \gm_2(\phi_y)\inv$, and all 
$\Gamma_{l},\, l\! \geq \!3$, are then determined recursively.
The direct recursion turns out to become intractable beyond $O(\kappa^6)$,
say. However, a closed graph theoretical solution of the recursion
can be obtained that yields $\Gamma_l$ for any $l \geq 1$
\cite{GraphJMP}. Importantly, the series in (\ref{i3}) can be expected
to have finite radius of convergence $\kappa < \kappa_c$,
at least as far as the associated vertex functions are concerned \cite{ReiszLRH}.
Once the series (\ref{i3}) has been constructed, an in principle exact solution
of (\ref{FRG1})'s lattice counterpart arises simply by substitution, $\Gamma_k =
\Gamma_{\kappa}|_{\ell \mapsto \ell(k)}$, for suitable $\ell(k)$ playing the role
of $R_k$. This differs from the standard uses of the FRG \cite{FRGbooks} in the
way initial conditions are imposed: $R_k$ is chosen such that $\Gamma_{k_0}[\phi]$,
for some finite $k=k_0$, is (up to kinematical factors) determined by the
above $\Gamma_0[\phi] = \sum_x \gamma(\phi_x)$. In overview, we propose to use the graph rule \cite{GraphJMP} for the computation
of vertex (and other correlation) functions but determine
bulk quantities from the FRGs (\ref{FRG1}), (\ref{i3}).

\section{Critical behavior from the LPA's unstable manifold}

For the hopping expansion the locus
of infinite correlation length (approached from the symmetric phase) is deemed to coincide with the radius
of convergence $\kappa_c$ of the (two-point and then all other) susceptibilities.
Traditionally, $\kappa_c$  has been estimated by pushing their
hopping expansion to high orders, at considerable effort. Our
proposed alternative rests on two simple observations: (i) any bulk
quantity other than a susceptibility should give the same $\kappa_c$,
in particular $\Gamma_{\kappa}$ for constant field (identified with the lattice average) should be
a legitimate choice. (ii) For $\kappa < \kappa_c$, specialization to
constant fields and resummation in (\ref{i3}) are commuting operations. 
The FRG (\ref{FRG1}) specialized to constant fields (up to kinematical factors) is known
as the Local Potential Approximation (LPA), or its modified version LPA' (see first ref. in \cite{FRGbooks}).  On general grounds, the relation between the
bare and renormalized parameters can be found by injecting initial data (determined by the
bare parameters) at the ultralocal scale $k=k_0$, and running the flow equation
for (truncations of) $\Gamma_k$  to the fixed point at $k \approx0$. This yields the
correlated values of the bare parameters in the action tuned to ensure that
$\Gamma[\phi]$ is based on the fixed point, i.e.~the unstable manifold of the
fixed point in question. In summary, one should be able to determine $\kappa_c$
from the unstable manifold of the LPA (or LPA') approximation to the FRG (\ref{FRG1}).

Explicitly, the following LPA ansatz $\Gamma_k[\phi_0]= a^D
\sum_x \{\! - \frac{1}{2}  \phi_0(x) (\Delta \phi_0)(x) 
+ U_k(\phi_0(x))\}$, is taken as the starting point, where $\Delta$ is the
lattice Laplacian and $a$ the lattice spacing. The flow equation (\ref{FRG1})
then specializes to
\begin{equation}
\label{LPA1}
\dd_k U_k(\phi_0) = \frac{1}{2} \int_{-\pi/a}^{\pi/a} 
\! \frac{d^Dp}{(2 \pi)^D} \, \frac{ \dd_k R_k(p)}{
\hat{p}^2 + R_k(p) + 
	U''_k(\phi_0)}\,,
\end{equation}
where the $k$ differentiation is at fixed $\phi_0$ and 
$\hat{p}^2 = \frac{4}{a^2} \sum_j \sin^2 \frac{p_ja}{2}$. 
Next we fix lattice units ($a=1$) and choose $R_k(p)$ 
to be the lattice step function $R_k(p) = (k^2 - \hat{p}^2) 
\th(k^2 - \hat{p}^2)$. Then  
\begin{equation}
\label{LPA2} 
k \dd_k U_k(\phi_0) =  \frac{k^2 \,{\rm Vol}(k)}%
{k^2 + U''_k(\phi_0) }\,, 
\quad {\rm Vol}(k) := \int_{-\pi}^{\pi} \frac{dp}{(2 \pi)^D} 
\,\th(k^2\! - \!\hat{p}^2)\,.
\end{equation}

The above flow equations and the fields have been 
dimensionful, and as such they do not lend 
themselves to a fixed point analysis. In particular the dimensionful LPA potential $U_k(\phi_0)$ will not itself reach a fixed point, but rather exhibit a characteristic scaling behavior. To proceed  we transition to a 
dimensionless LPA formulation by rescaling both the field and potential,
\begin{equation} 
\label{lpa1} 
V_k(\phi) := \frac{1}{\mu_D k^D} U_k(\phi_0(\phi)) \,, 
\quad \phi_0(\phi) := k^{\frac{D-2}{2}} 
\sqrt{\mu_D} \phi\,,\quad \mu_D := \frac{1}{ (4\pi)^{D/2} 
	\Gamma(\frac{D}{2} +1) }\,.
\end{equation}
The new potential $V_k(\phi)$ is a dimensionless function of the 
dimensionless field $\phi$. 
The constants are adjusted such that 
only the dimensionless volume function ${\rm vol}(s) := {\rm Vol}(k_0 s)/(\mu_D k_0^Ds^D)$, $0 \leq s := k/k_0 \leq 1$, $k_0:=\sqrt{4D}$,
occurs. It is determined numerically, and is roughly bell shaped with a maximum of $O(1)$. The normalizations are such that ${\rm vol}(0) =1$ and  $ {\rm vol}(1) 
= 1/\mu_D k_0^D$. 

 The dimensionful flow 
equation (\ref{LPA2}) translates into 
\begin{equation} 
\label{lpa3} 
s \dd_s V_s (\phi) = - D V_s(\phi) + 
\frac{D\!-\!2}{2} \phi V_s'(\phi) + 
\frac{ {\rm vol}(s)}{1 + V_s''(\phi)}\,.
\end{equation}
In this form one can now search 
meaningfully for a $s \ra 0$ fixed point potential solving 
\begin{equation}  
\label{lpa4}
0 = - D V_*(\phi) + 
\frac{D\!-\!2}{2} \phi V_*'(\phi) + 
\frac{1}{1 + V_*''(\phi)}\,.
\end{equation}
To proceed,  insertion of the Taylor series $V_s(\phi) = \sum_{i \geq 0} \frac{g_{2i}(s)}{(2 i)!}\ph^{2i} $ into (\ref{lpa3}) produces the beta functions for the couplings $g_{2i}(s)$,
\begin{eqnarray}
\label{lpa6} 
s \dd_s g_{2i} \is \beta_{2i}( g_2,\ldots, g_{2 i + 2}) \,,\quad i \geq 1\,,
\nonum
\beta_2 \is -2 g_2 - \frac{{\rm vol}(s)}{ (1 + g_2)^2} \,g_4\,,\quad \beta_4 = (D\!-\!4) g_4  - \frac{{\rm vol}(s)}{ (1 + g_2)^2} \,g_6
+\frac{6{\rm vol}(s)}{ (1 + g_2)^3} \,g_4^2\,,
\end{eqnarray}
etc.~Truncating via $g_{2 N+2} \equiv 0$ at some order $N$ 
a closed system of $N+1$ ODEs arises. The lowest order equation $s \dd_s g_0 = 
- D g_0 + {\rm vol}(s)/(1+ g_2)$ determines $g_0$ once $g_2$ 
is known.

Clearly, the behavior of the dimensionless couplings 
in the vicinity of the fixed point is instrumental
for the critical behavior. The fixed couplings themselves
obey a simple recursion relation of the form $g^*_{2i} = - 2 g_2^* (1 + g_2^*)^i P_{i-2}(g_2^*)$,
where generally $P_n$ is a polynomial of degree $n$. The
value of $g_2^*$ is constrained by the truncation condition 
$g^*_{2N+2} =0$, \ie it is the root of the high order polynomial 
equation $ 2 g_2^* (1 + g_2^*)^{n+1} P_{n-1}(g_2^*)=0$. In general this equation produces many spurious `fixed points', but  the `correct' solution can be selected by requiring that its corresponding stability matrix,
\begin{eqnarray}
\label{stabmat1}
M(g^*)_{ij} := \frac{\dd \beta_{2i}}{\dd g_{2j}} \bigg|_{g = g^*}\,,
\end{eqnarray}
 have precisely one negative eigenvalue, conventionally denoted by $-\theta_1$. For $D\geq 3$ one always has a Gaussian fixed point with $-\theta_1=-2$. In $D=3$ one finds in addition the Fisher-Wilson fixed point with $-\theta_1\approx-1.5396$.

The stability matrix $M$ determines the flow pattern in the vicinity of the fixed point. Writing $\delta g = (\delta g_2, \delta g_4, 
\ldots, \delta g_{2N})$ for a perturbation about the  fixed point $g^*$ and expanding the 
truncated system (\ref{lpa6})  to linear order in the perturbation gives $s \dd_s \delta g = M(g^*) \delta g$. The solution of this linearized flow can be written in terms of the eigensystem $(v^{(j)},-\theta_j)$ of $M$ as 
\begin{eqnarray}
\label{lpa11}
\delta g(s)= \sum_{j=1}^N c_j v^{(j)} s^{-\theta_j}\,,
\end{eqnarray}
where the boundary constants $c_1,\ldots,c_N$ are set at some scale $0<s_0\ll 1$. For small $s$ and $c_2 \neq 0$ 
the sum will be dominated by the $j=1$ term (as $-\theta_1<0$) and thus blow up, taking the couplings away from the fixed point. 
Conversely, the locus of linearized couplings that flow to $g^*$ 
for $s \ra 0$ is characterized by $c_2 =0$. More intrinsically, 
there exists a unique linear combination such that 
\begin{equation}
\label{lunstab2} 
a_0 + \sum_{i=1}^N a_i \,g_{2i} (s) = {\rm const} \,c_2 \,
s^{-\theta_1}\,,\quad 0< s \leq s_0 \ll 1\,.
\end{equation}
For $c_2 =0$ this linear combination describes the 
``linearized unstable manifold'', i.e.~the codimension one 
hyperplane from which the couplings flow into the fixed point. The coefficients of the unstable manifold can be computed analytically for the Gaussian fixed point in $D=4$, and come out as $a_0= 0$, $a_i=1/[2^{i-1}(i-1)!]$, for $1\leq i\leq N$.

So far no action-specific information has entered. The fixed point,
the critical exponents, and the linearized unstable manifold are  
computable solely in terms of the field content (here: one scalar 
field), the dimensionality  $D$, the nature 
of the truncation (here: LPA), and the mode modulator 
(here: the lattice step function). Action-specific information 
is in the present setting injected by specifying initial data $g_{2i}(s=1)$, 
computed from the ultralocal part of the theory's action. 

In order to obtain this intial data, we first note that at any scale $k$, the Legendre effective action $\Gamma_k$ in (\ref{FRG1}) satisfies the functional integro-differential equation
\begin{eqnarray}
\label{ul1}
e^{-\Gamma_k[\ph_0]}=\int\! \prod_{x} d\chi_0(x) e^{-S[\chi_0]-\frac{1}{2}(\ph_0-\chi_0)\cdot R_k \cdot (\ph_0-\chi_0)-(\ph_0-\chi_0)\cdot \frac{\delta \Gamma_k}{\delta \ph_0}
}\,,
\end{eqnarray}
where $S[\chi_0]=\frac{1}{2}\sum_{x}\big\{ -\chi_0(x)(\Delta \chi_0)(x)+m_0^2\chi_0(x)^2+\lb_0 \chi_0(x)^4/12\big\}$ is the bare action, and `$\cdot$' denotes a sum over lattice sites. Although this equation cannot in general be solved exactly, at the ultralocal scale $k_0=\sqrt{4D}$ where $R_{k_0}(x,y)=k_0^2\delta_{xy}+\Delta_{xy}$, (\ref{ul1}) transcribes to 
\begin{eqnarray}
\label{ul2}
e^{-\sum_x U_{k_0}(\ph_0(x))}=\int\! \prod_{x} d\chi_0(x) e^{-S_0[\chi_0]-\frac{k_0^2}{2}(\ph_0-\chi_0)\cdot (\ph_0-\chi_0)-(\ph_0-\chi_0)\cdot U_{k_0}'
}\,.
\end{eqnarray}
Here $S_0[\chi_0]$ is the ultralocal part of the bare action, and $U_{k_0}$ is the effective potential at $k=k_0$. Since all the quantities in (\ref{ul2}) are ultralocal, the functional integral factorizes. After transitioning to a dimensionless potential $V_s(\ph)$ viz. (\ref{lpa1}), and hopping parametrizing the bare action, we obtain the ordinary integro-differential equation satisfied by $V_{s=1}(\ph)$
\begin{eqnarray}
\label{ul3}
e^{-{\rm vol}(1)\inv V_{s=1}(\ph)}=\int \! d\chi e^{-\frac{1-2\lb-D\kappa}{4D{\rm vol}(1)\kappa}\chi^2-\frac{\lb}{(4D{\rm vol}(1)\kappa)^2}\chi^4-\frac{1}{2{\rm vol}(1)}(\ph-\chi)^2-{\rm vol}(1)\inv (\ph-\chi)V_{s=1}'(\ph)}\,.
\end{eqnarray}
Inserting the truncation ansatz $V_{s=1}(\phi) = \sum_{i = 0}^N \frac{g_{2i}(s=1)}{(2 i)!}\ph^{2i} $ into (\ref{ul3}) then determines the initial couplings $g_{2i}(s=1)$ as functions of the bare parameters $\kappa,\,\lb$ via exactly computable single site integrals. 

With the initial data  known, the integration of the flow equations 
(\ref{lpa6}) (truncated at some order $N$) 
proceeds as follows. Since the initial data $g_{2i}(s=1)$, $1 \leq i \leq N$, 
are prescribed functions of $\kappa, \lb$, a well-defined evolution 
via the ODE system will render the $g_{2i}(s) = g_{2i}(s| \kappa,\lb)$
parametrically dependent on $\kappa, \lb$ for all $s$ for which 
the evolution is regular. For generic $\kappa,\lb$ the flow will 
{\it not} come close to the fixed point; it will do so however 
once it reaches the linearized unstable manifold at some $0 < s_0 \ll 1$.
In a given polynomial approximation of order $N$ one therefore needs to 
solve
\begin{equation}
\label{shooting1} 
a_0+\sum_{i=1}^N a_i g_{2i}(s_0|\kappa,\lb) =0 \quad \Longrightarrow  
\quad \kappa = \kappa_c(\lb)\,.
\end{equation}
In the present context this will ensure that the 
flow very nearly reaches the fixed point, with limitations only 
set by numerical accuracy. 

\section{Results for $\phi_3^4$ and $\phi_4^4$} 

The above technique has been applied to determine $\kappa_c(\lb)$ for
$\phi^4$ theories in both $D=3$ and $D=4$. The shooting technique has
been implemented in {\tt Mathematica} without encountering
significant obstructions from stiffness for reasonably large $N$.
Related results have been obtained in \cite{Caillold3}, \cite{Caillold4} in a different LPA
formulation but without relation to (\ref{i3}) and the hopping expansion's
radius of convergence.  

As a proof of principle  we first applied the shooting technique in $D\!=\!3$,  aiming at the Fisher-Wilson fixed point. Since in $D=3$ the anomalous dimension $\eta$ is non-zero, the neglection of a wavefunction renormalization constant in the LPA (as opposed to the LPA' ansatz) induces a systematic error. Nevertheless, the comparison of the LPA results with Monte-Carlo data \cite{Hasenbusch} shows reasonable agreement.

\begin{table}[h]
	\begin{center}
		\begin{tabular}{|c|c| c||c|c| c|}
			\hline
			$\lambda$&$\kappa_{c,MC}$ & $\kappa_c$&$\lambda$&$\kappa_{c,MC}$ & $\kappa_c$\\
			\hline
			0.1& 0.37341&0.3732&0.9&0.38451 & 0.3854\\
			0.2& 0.3884&0.3882&1.3& 0.36522& 0.3659\\
			0.4& 0.3975& 0.3975&1.4&  0.36028& 0.362\\
			0.7& 0.39253& 0.3926&1.5&0.3553 & 0.358\\
			0.8& 0.3887&0.3898&	2.5&0.3134 &0.3149\\
			\hline
		\end{tabular}
	\end{center}
	\caption{Critical values for $\phi_3^4$ theory in $D=3$. Left, $\kappa_{c,MC}$ from 
		\cite{Hasenbusch} with only significant digits displayed. Right $\kappa_c$ from LPA at truncation order $N=20$. The LPA errors are a combination of numerical and estimated truncation effects, only significant digits are displayed. The discrepancy can plausibly be attributed to the neglected anomalous dimension.} 
\end{table}


In $D=4$ only the Gaussian fixed point is found. By working with the LPA the $\eta=0$ contention \cite{rosten} is probed for self consistency. As an illustration of the shooting technique we depict in Figs.~1(a) and 1(b) the flow of 
the couplings $g_{2}(s)$, $g_{4}(s)$, $g_{6}(s)$, $g_{8}(s)$, and $g_{10}(s)$ in the $\phi_4^4$
theory towards the Gaussian fixed point.
The truncation order is $N=20$, with $\kappa$ adjusted at fixed $\lb$ such that the
$g_{2i}(s_0=0.001|\kappa,\lb)$ satisfy (\ref{shooting1}), with coefficients $a_0= 0$, $a_i=1/[2^{i-1}(i-1)!]$, for $1\leq i\leq N$.

\begin{figure}[h]
	\centering
	\begin{subfigure}{.4\textwidth}
		\centering
			\includegraphics[scale=0.35]{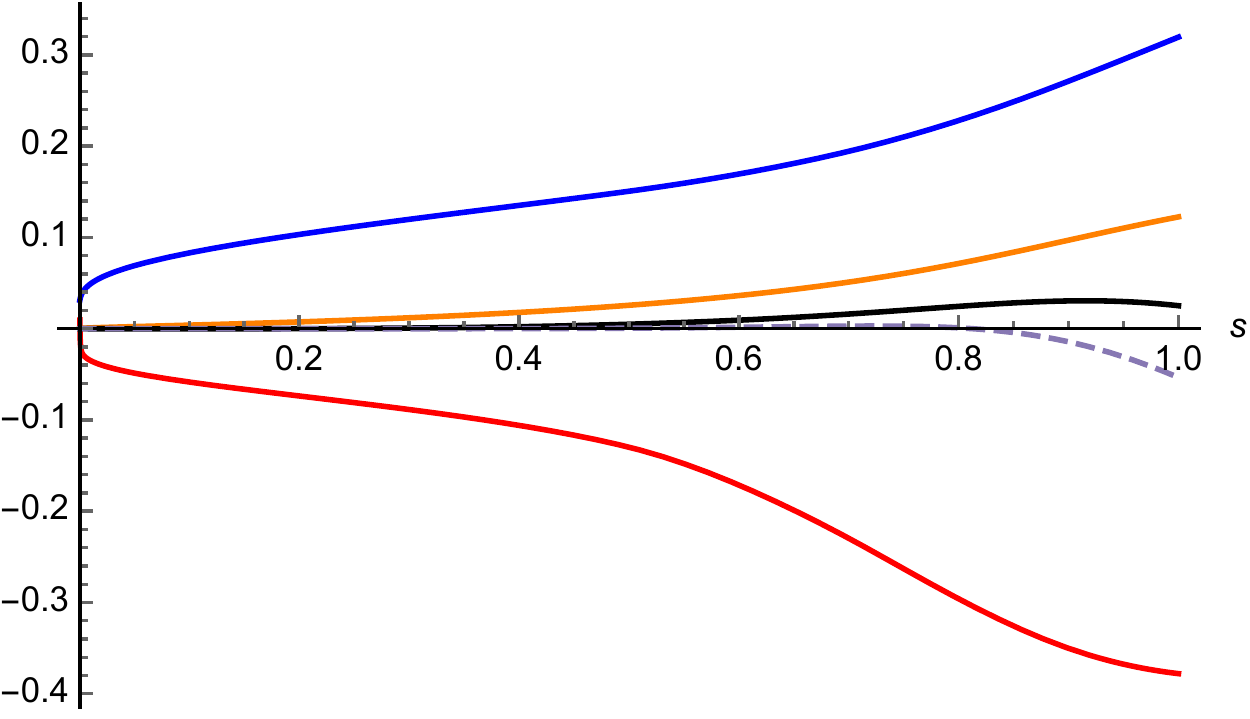}
		\caption{}
	\end{subfigure}
	\begin{subfigure}{.4\textwidth}
		\centering
		\includegraphics[scale=0.35]{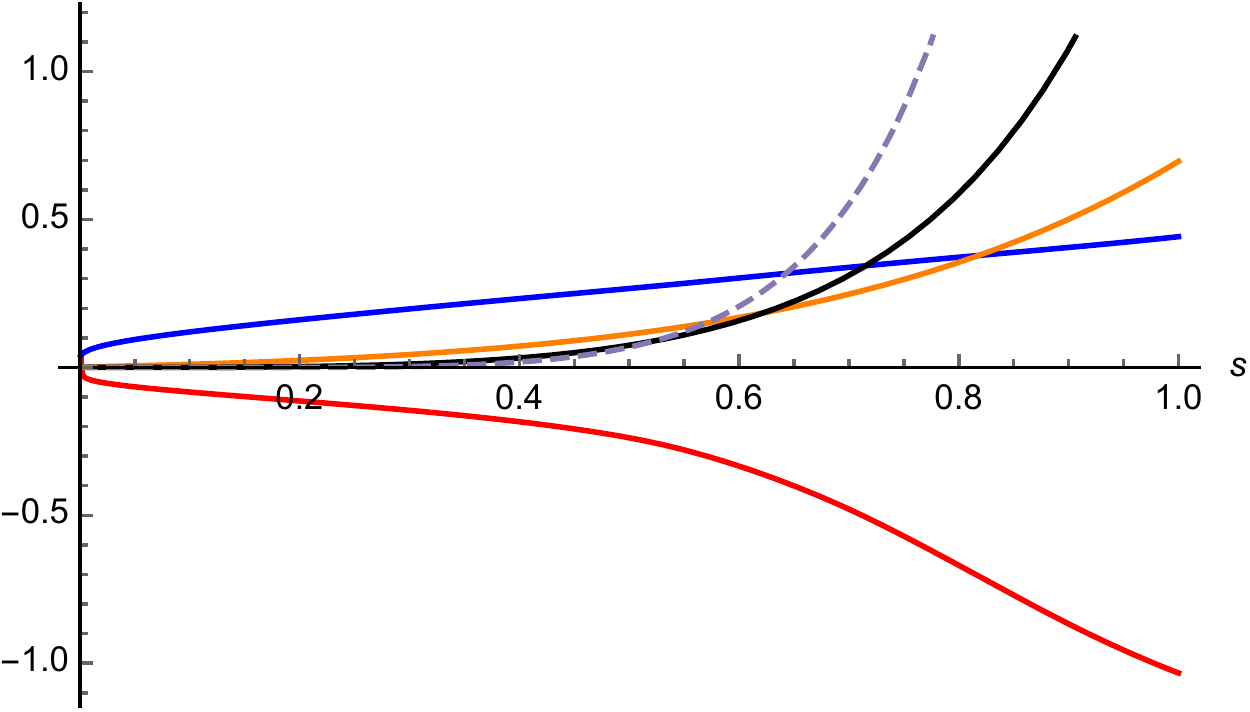}
		\caption{}
	\end{subfigure}
	\caption{$\phi_4^4$: flow of couplings $g_{2}(s), \,g_{4}(s),\,g_{6}(s),\,g_{8}(s),\,g_{10}(s)$ 
		for (a) $(\lambda,\kappa)=(0.48548, 0.2828)$, and (b) $(\lambda,\kappa)=(4.3303, 0.1834)$.  
		\texttt{Red}:  $g_{2}$, \texttt{Blue}:  $g_{4}$, \texttt{Orange}:  
		$g_{6}$,  \texttt{Black}:  $g_{8}$, \texttt{Dashed}:  $g_{10}$. }
\end{figure}

As noted earlier, the critical line $\kappa_c(\lb)$ has been previously computed 
from the radius of convergence of the hopping expansion in \cite{Luscher1}. 
A comparison of our results with the $\kappa_c(\lb)$ values of L\"{u}scher-Weisz 
(taken from Table 1 in \cite{Luscher1})  is presented in Table 2.

\begin{table}[h]
	\begin{center}
		\begin{tabular}{|c| c|  c| }
			\hline	
			$\lb$&$\kappa_{c,LW}$&$\kappa_c/2$ \\
			\hline
			0&0.1250(1)&0.1250(1)\\
			2.4841$\times 10^{-2}$&0.1294(1)&0.12928(3)\\
			3.5562$\times 10^{-2}$&0.1308(1)&0.13068(3)\\
			1.3418$\times 10^{-1}$&0.1385(1)&0.1381(4)\\
			2.7538$\times 10^{-1}$&0.1421(1)&0.1416(4)\\
			4.8548$\times 10^{-1}$&0.1418(1)&0.1414(4)\\
			7.7841$\times 10^{-1}$&0.1376(1)&0.1374(4)\\
			1.7320&0.1194(1)&0.1190(5)\\
			2.5836&0.1067(1)&0.1066(5)\\
			4.3303&0.09220(9)&0.0917(7)\\
			$\infty$ (LW) or 100 (LPA) &0.07475(7)&0.0722(1)\\
			\hline
		\end{tabular}
	\end{center}
	\caption{Critical values for $\phi_4^4$ theory in $D=4$. Left, $\kappa_{c,LW}$ from 
		L\"{u}scher-Weisz \cite{Luscher1}. Right $\kappa_c/2$ from LPA at truncation order $N=20$.  The errors in the LPA results are a combination of numerical and estimated truncation errors.}
	\vspace{-4mm}
\end{table}

In summary, the critical behavior in the hopping expansion, traditionally set by the radius of convergence \cite{Luscher1}, \cite{ReiszLRH}, can alternatively be obtained simply from the LPA or LPA' approximation to the FRG.

\medskip

{\it {\bf Acknowledgements:}  Support by a PITTPACC fellowship, and University of Pittsburgh A\&S-PBC, A\&S-GSO travel grants  is gratefully acknowledged. With many thanks to Max Niedermaier. }

\end{document}